\documentclass[11pt,a4j]{article}
\usepackage{amsmath}
\usepackage{amsthm}
\usepackage{overcite}
\newtheorem{proposition}{Proposition}
\newtheorem{definition}{Definition}
\newtheorem{theorem}{Theorem}
\newtheorem{corollary}{Corollary}

\def\C{\mathbf{C}}
\begin{document}

\begin{center}
  {\LARGE \textbf{New Integrable Family in the $n$-Dimensional
  Homogeneous Lotka--Volterra Systems\\
  with Abelian Lie Algebra\\
  }}

\vskip 1cm

{\large Kenji \textsc{Imai}$^{\dag}$ and Yoshihiro \textsc{Hirata}$^{\ddag}$}
{}~\\
\quad \\
{\em ~$~^{\dag}$School of Liberal Arts and Sciences, 
  Daido Institute of Technology, }\\
{\em Nagoya, 457-8530, Japan}\\
 {\tt kimai@daido-it.ac.jp}
{}~\\
\quad \\
{\em ~$~^{\ddag}$Graduate School of Human Informatics, 
  Nagoya University,}\\
{\em Nagoya, 464-8601, Japan}\\
{\tt hirata@ncube.human.nagoya-u.ac.jp}

\end{center}

\vskip 1cm {\bf Abstract.} {\small 
  We present an $n$-dimensional integrable homogeneous Lotka--Volterra
  system, which has $(n^2-1)$-dimensional Lie symmetry algebra.
  Moreover a wider integrable family is derived from the structure of
  the Lie algebra.
}

\vskip 1cm

The $n$-dimensional homogeneous Lotka--Volterra (HLV) system is given
by
\begin{equation}
  \label{eq:HLV}
  \frac{\mathrm{d}x_i}{\mathrm{d}t} = f_i(x) = x_i \sum_{j=1}^{n}
  a_{ij} x_j, \quad i = 1, \dots, n,
\end{equation}
where $a_{ij}, \, i,j = 1, \dots, n$ are complex parameters.
Integrability of the HLV system~\eqref{eq:HLV} has been fairly
investigated in some particular cases, \textit{e.g.} soliton
systems\cite{HS76,Nar82,Ito87}, the $2$-dimensional
systems\cite{CFL99,CL00b} or the $3$-dimensional ABC
system\cite{AMM95,Lab96,Mou97,Gao99,CL00a}.
Almost all the above studies dealt with first integrals in the
criterions of integrability.
However it is known that a system of ordinary differential equations
is also integrable if the system has enough Lie symmetry vector
fields.
In this letter we treat Lie symmetries and find a new integrable
family in the $n$-dimensional HLV systems, named the ladder system or
the generalized ladder system.


We rewrite the considered system~\eqref{eq:HLV} into a vector field
form as
\begin{equation*}
  \label{eq:VecFld}
  X_f = \sum_{i=1}^{n} f_i(x) \frac{\partial}{\partial x_i}.
\end{equation*}
A Lie symmetry is a vector field which commutes the vector field
$X_f$.
\begin{definition}[Lie symmetry]
  A vector field $X$ is called a Lie symmetry with respect to the
  system~\eqref{eq:HLV} if $X$ commutes $X_f$, i.e.,
  \begin{equation*}
    \label{eq:commute}
    [X_f, X] := X_f X - X X_f = 0.
  \end{equation*}
\end{definition}
A set of all the Lie symmetries forms a Lie algebra, and if the
algebra has an $(n-1)$-dimensional solvable sub-algebra, the system
can be integrated by quadrature.

Let us consider a particular system in the HLV systems, which we call
the $n$-dimensional ladder system.
We introduced the $3$-dimensional ladder system in Ref.\,[\citen{HI02}].
The $n$-dimensional ladder system possesses polynomial Lie symmetries
which constitute $(n-1)$-dimensional Abelian Lie algebra.
Hence the $n$-dimensional ladder system can be integrated by using the
Lie algebra.
\begin{definition}[The ladder system]
  The $n$-dimensional HLV system~\eqref{eq:HLV} with the coefficients
  \begin{equation}
    \label{eq:matrixL}
    A := (a_{ij})_{1 \le i,j \le n} =
    \left(
      \begin{array}{cccc}
        1 & 0 & \cdots & -n+2 \\
        2 & 1 & \cdots & -n+3 \\
        \vdots & \vdots & \ddots & \vdots \\
        n & n-1 & \cdots & 1
      \end{array}
    \right)
  \end{equation}
  is called the $n$-dimensional (homogeneous Lotka--Volterra) ladder
  system.
\end{definition}
The ladder system has rational Lie symmetries as introduced below.
\begin{theorem}
  \label{th:HLVL}
  The $n$-dimensional ladder system~\eqref{eq:HLV} with
  eq.~\eqref{eq:matrixL} possesses the following Lie symmetries:
  \begin{equation*}
    \label{eq:SymHLVL}
    Y^l_m = x_m u^{l-m-1}
    \left(
      D - u \frac{\partial}{\partial x_l}
    \right), \quad l,m = 1, \dots, n,
  \end{equation*}
  where
  \begin{eqnarray*}
    u = \sum_{j=1}^n x_j, \\
    D = \sum_{j=1}^n x_j \frac{\partial}{\partial x_j}.
  \end{eqnarray*}
\end{theorem}
A proof is straightforward and hence we omit it.
Now we also present several basic properties for $Y_m^l$ without
proofs.
\begin{proposition}
  \begin{gather}
    \label{eq:commutator}
    [ Y_m^l, Y_{m^\prime}^{l^\prime} ] = \delta_{m^\prime, l}
    Y_m^{l^\prime} - \delta_{m, l^\prime} Y_{m^\prime}^l, \\
    \label{eq:Y_l^l}
    \sum_{l=1}^{n} Y_l^l = 0. \nonumber \\
    \label{eq:Ymlu}
    Y_m^l (u) = 0, \\
    \label{eq:YmlD}
    [ Y_m^l, D ] = ( l - m ) Y_m^l, \nonumber
  \end{gather}
  and especially,
  \begin{equation*}
    \label{eq:YllD}
    [ Y_l^l, D ] = 0.
  \end{equation*}
\end{proposition}
Now let us introduce the following linear space
\begin{equation*}
  \label{eq:L}
  \mathcal{L} =
  \left\{
    \biggl.
    \sum_{l,m = 1}^{n} a_l^m Y_m^l
    \biggr|
    a_l^m \in \C
  \right\}.
\end{equation*}
\begin{theorem}
  \label{th:L}
  The linear space $\mathcal{L}$ forms $(n^2-1)$-dimensional Lie
  algebra.
\end{theorem}
\begin{proof}
  It is evident that $\mathcal{L}$ forms a closed algebra from
  eq.~\eqref{eq:commutator}.
  Hence we only show that $\mathcal{L}$ has $n^2-1$ dimensions.

  $Y_m^l$ have homogeneous rational coefficients of order $l-m$.
  We therefore consider the dimension of each homogeneous subset.
  First, consider the linear combination of $Y_{l-k}^l, \, -n < k < n,
  \, k \ne 0$ as
  \begin{equation*}
    \label{eq:lincom}
    \sum_{l = k+1}^{k+n} \alpha_l Y_{l-k}^l = 0,
  \end{equation*}
  where $\alpha_l \in \C$ and $\alpha_l = 0$ for $l \le 0, \, n + 1
  \le l$.
  If $k > 0$, the coefficient of $\partial / \partial x_1$ is written
  as
  \begin{equation*}
    \label{eq:coeff1}
    u^{k-1} \sum_{l=k+1}^{k+n} \alpha_l x_{l-k} x_1.
  \end{equation*}
  Obviously $\forall l$, $\alpha_l = 0$.
  Hence when $k > 0$, $Y_{l-k}^l$ are linearly independent over $\C$.
  In the same way, if $k < 0$, by considering the coefficient of
  $\partial / \partial x_n$, one can show the linear independency.

  Next we deal with the ``diagonal'' elements $Y_l^l$.
  The linear combination of $Y_l^l, \, 1 \le l \le n$ is rewritten as
  \begin{align*}
    \label{eq:k=0}
    \sum_{l=1}^n \alpha_l Y_l^l &= u^{-1}
    \left(
      \sum_{l=1}^n \alpha_l x_l \sum_{j=1}^n x_j
      \frac{\partial}{\partial x_j} - \sum_{l=1}^n \alpha_l x_l
      \sum_{j=1}^n x_j \frac{\partial}{\partial x_l}
    \right) \nonumber \\
    &= u^{-1}
    \left(
      \sum_{j=1}^n \alpha_j x_j \sum_{l=1}^n x_l
      \frac{\partial}{\partial x_l} - \sum_{l=1}^n \alpha_l x_l
      \sum_{j=1}^n x_j \frac{\partial}{\partial x_l}
    \right) \nonumber \\
    &= u^{-1} \sum_{j=1}^n \sum_{j=1}^n ( \alpha_j - \alpha_l ) x_j x_l
    \frac{\partial}{\partial x_l}.
  \end{align*}
  Hence $\sum \alpha_l Y_l^l = 0$ if and only if $\alpha_1 = \dots =
  \alpha_n$.
  Thus the linear space
  \begin{equation*}
    \label{eq:Lp}
    \mathcal{L}^\prime =
    \left\{
      \biggl.
        \sum_{l=1}^{n} a_l Y_l^l
      \right|
      a_l \in \C
    \biggr\}
  \end{equation*}
  has $n-1$ dimensions.
  This completes a proof.
\end{proof}
In particular $\mathcal{L}$ has $(n-1)$-dimensional Abelian
sub-algebra constituted with polynomial Lie symmetries.
\begin{corollary}
  The polynomial $(n-1)$-dimensional sub-algebra spanned by $\{ Y_1^n,
  \dots, Y_{n-1}^n \}$ is Abelian.
  Hence the ladder system is integrable.
\end{corollary}
The following statement, obtained as a by-product of the proof of
Theorem~\ref{th:L}, is essential for generalization of the ladder
system.
\begin{corollary}
  \label{th:diagonal}
  The $(n-1)$-dimensional linear space
  $\mathcal{L}^\prime$ forms $(n-1)$-dimensional Abelian sub-algebra.
\end{corollary}

Now we generalize the ladder system by using the Abelian sub-algebra
$\mathcal{L}^\prime$.
Since $[ Y_l^l, Y_m^m ] = 0$ and $[ Y_l^l, D ] =0$, arbitrary linear
combinations of $Y_m^m, \, m = 1, \dots, n$ and $D$ over $\C$ commute
every $Y_l^l$ as
\begin{equation*}
  \label{eq:mainpre1}
  \Bigl[
    Y_l^l, \alpha_0 D - \sum_{m=1}^n \alpha_m Y_m^m
  \Bigr] = 0,
\end{equation*}
where $\alpha_m \in \C, \, m = 0, 1, \dots, n$.
Moreover from eq.~\eqref{eq:Ymlu}, the following relation
\begin{equation}
  \label{eq:mainpre2}
  \Bigl[
    Y_l^l, \alpha_0 u D - u\sum_{m=1}^n \alpha_m Y_m^m
  \Bigr] = 0
\end{equation}
holds.

On the other hand
\begin{align*}
  \alpha_0 uD - u\sum_{m=1}^n \alpha_m Y_m^m &= \alpha_0 u
  \sum_{j=1}^n x_j \frac{\partial}{\partial x_j} - \sum_{m=1}^n
  \alpha_m x_m \left(
    D - u \frac{\partial}{\partial x_m}
  \right) \\
  &= \alpha_0 \sum_{i,j = 1}^n x_i x_j \frac{\partial}{\partial x_j}
  - \sum_{m=1}^n \alpha_m x_m \sum_{k=1}^n x_k
  \left(
    \frac{\partial}{\partial x_k} - \frac{\partial}{\partial x_m}
  \right) \\
  &= \alpha_0 \sum_{i,j = 1}^n x_i x_j \frac{\partial}{\partial x_i}
  - \sum_{k,m = 1}^n \alpha_m x_k x_m \frac{\partial}{\partial x_k} +
  \sum_{k,m = 1}^n \alpha_m x_k x_m \frac{\partial}{\partial x_m} \\
  \label{eq:mainpre3}
  &= \sum_{i,j = 1}^n ( \alpha_0 + \alpha_i - \alpha_j ) x_i x_j
  \frac{\partial}{\partial x_i}.
\end{align*}
Hence $\alpha_0 uD - u\sum_{m=1}^n \alpha_m Y_m^m$ defines the HLV
system with the coefficients $a_{ij} = \alpha_0 + \alpha_i -
\alpha_j$.
It can be regarded as a generalization of the ladder system: if one
puts $\alpha_0 = 1, \, \alpha_j = j, \, j = 1, \dots, n$, then the
generalized ladder system becomes the ladder system.
\begin{definition}[The generalized ladder system]
  The HLV system~\eqref{eq:HLV} with $a_{ij} = \alpha_0 + \alpha_i -
  \alpha_j, \, \alpha_j \in \C, \, 0 \le j \le n$ is called the
  (homogeneous Lotka--Volterra) generalized ladder system.
\end{definition}
Equation~\eqref{eq:mainpre2} implies the following statement.
\begin{theorem}
  The generalized ladder system possesses the $(n-1)$-dimensional
  Abelian Lie algebra $\mathcal{L}^\prime$.
  Hence it is integrable.
\end{theorem}

The generalized ladder system also possesses the following Lie
symmetries besides $Y_l^l, \, l = 1, \dots, n$.
\begin{proposition}
  \label{th:Ytilde}
  $\tilde{Y}_m^l = x_m u^{\alpha_l - \alpha_m - 1} ( D - u \partial /
  \partial x_l ), \, l,m = 1, \dots, n$ are Lie symmetries for the
  generalized ladder system.
\end{proposition}
A proof is straightforward.
Note that the above Lie symmetries are logarithmic in general.
Moreover the $(n^2 - 1)$-dimensional linear space
$\tilde{\mathcal{L}}$ defined by
\begin{equation*}
  \label{eq:Ltilde}
  \tilde{\mathcal{L}} =
  \left\{
    \biggl.
      \sum_{m,l = 1}^n a_l^m \tilde{Y}_m^l
    \biggr|
    a_l^m \in \C
  \right\}
\end{equation*}
has the same algebraic structure with $\mathcal{L}$.
One can easily show the following proposition.
\begin{proposition}
  The commutators among $\tilde{Y}_m^l, \, l,m = 1, \dots, n$ are
  given by
  \begin{equation*}
    \label{eq:com2}
    [ \tilde{Y}_m^l, \tilde{Y}_{m^\prime}^{l^\prime} ] =
    \delta_{m^\prime, l} \tilde{Y}_m^{l^\prime} - \delta_{m, l^\prime}
    \tilde{Y}_{m^\prime}^l.
  \end{equation*}
  Hence $\tilde{\mathcal{L}} \cong \mathcal{L}$.
\end{proposition}
Thus although we construct the generalized ladder system by extracting
the Abelian sub-algebra $\mathcal{L}^\prime$, the same algebraic
structure is restored.

Since the generalized ladder system is integrable, there exist $n-1$
first integrals.
However the first integrals are not always polynomial or rational
functions.
The first integrals can easily be composed by the Lie symmetries
$\tilde{Y}_m^l$.
\begin{proposition}
  $\displaystyle \frac{x_m}{x_l} u^{\alpha_l - \alpha_m}, l,m = 1,
  \dots, n$ are first integrals for the generalized ladder system.
\end{proposition}
\begin{proof}
  If there exist two Lie symmetries $X$ and $X^\prime$, and exists a
  function $F$ such that $X^\prime = F X$, then $F$ is a first
  integrals.
  One can compose the first integrals by using the obtained Lie
  symmetries in Proposition~\ref{th:Ytilde}.
\end{proof}

In this letter we have first introduced the $n$-dimensional ladder
system~\eqref{eq:HLV} with eq.~\eqref{eq:matrixL}.
The ladder system possesses the Lie symmetry algebra
$\mathcal{L}$, and it is integrable since $\mathcal{L}$ has
$(n-1)$-dimensional Abelian sub-algebra (Corollary~\ref{th:HLVL}).
Moreover, by extracting the Abelian Lie sub-algebra
$\mathcal{L}^\prime \subset \mathcal{L}$, we have constructed the new
integrable family in the HLV systems, the generalized ladder system.
Although $\mathcal{L}$ is not Lie symmetry algebra for the generalized
ladder system anymore, the generalized ladder system possesses the Lie
symmetry algebra $\tilde{\mathcal{L}}$, which is isomorphic to
$\mathcal{L}$.

The first integrals given in this letter are rational for the ladder
system and generally logarithmic for the generalized ladder system.
Thus despite the first integrals which belong to different classes, the
structure of the Lie symmetry algebra is just the same between the ladder and
the generalized ladder systems.


We have presented the method to derive new integrable systems using
Lie symmetry sub-algebra.
The universality of the method is open.

\end{document}